# Plasmonic waveguide-integrated nanowire laser

*Esteban Bermúdez-Ureña\*,1, Gozde Tutuncuoglu†2, Javier Cuerda†3, Cameron L. C. Smith4, Jorge Bravo-Abad3, Sergey I. Bozhevolnyi5, Anna Fontcuberta i Morral2, Francisco J. García-Vidal3,6, and Romain Quidant\*,1,7*

[1]ICFO-Institut de Ciencies Fotoniques, Barcelona Institute of Science and Technology, 08860 Castelldefels (Barcelona), Spain

[2]Laboratoire des Matériaux Semiconducteurs, École Polytechnique Fédérale de Lausanne, 1015 Lausanne, Switzerland

[3]Departamento de Física Teórica de la Materia Condensada and Condensed Matter Physics Center (IFIMAC), Universidad Autónoma de Madrid, 28049 Madrid, Spain

[4]Department of Micro- and Nanotechnology, Technical University of Denmark, DK-2800, Kongens Lyngby, Denmark

[5]Department of Technology and Innovation, University of Southern Denmark, Niels Bohr Allé 1, DK-5230 Odense M, Denmark

[6]Donostia International Physics Center (DIPC), E-20018 Donostia/San Sebastian, Spain

[7]ICREA–Institució Catalana de Recerca i Estudis Avançats, 08010 Barcelona, Spain

Integrated photonic circuits are at the heart of promising technological applications that range from optical interconnects for conventional data processing and communications[1,2] and quantum optics/information[3,4], to novel on-chip sensing platforms of chemical and biological species[5,6]. Ultimately, an ideal platform should integrate all key components within the same chip, including the light source, transmission lines, modulators and detectors[7]. Plasmonics, being concerned with the generation, control and detection of surface plasmon polariton (SPP) modes[8], can play a major role in the development of such integrated platforms[9,10]. In particular, plasmonic waveguides ensuring sub-diffraction field confinements at optical frequencies, are promising to couple together individual elements for usage in, for example, data centre technologies[11], as they can potentially enable hybrid photonic-plasmonic circuitry[12] with enhanced interactions, ultra-compact footprint, high bandwidth and low-power consumption[13].

Small on-chip lasers exhibiting dimensions or mode sizes comparable to or smaller than the emission wavelength represent an ideal solution for light source integration that may circumvent the need for free-space coupling mechanisms[14], and several candidates have emerged over the last couple of decades[15]. Plasmon lasers are especially promising due to their ability to confine the emission below the diffraction limit together with improved performances provided by the enhanced emission dynamics[16,17]. In this context, semiconductor nanowire (NW) lasers have received increasing interest owing to their intrinsic quasi-one-dimensional cavity geometry, the variety of high quality materials enabling lasing over a wide spectral range, from the near ultra-violet to the near infrared[18], and the potential to combine epitaxial heterostructures with unique optoelectronic properties[19].

In the technologically relevant near-infrared region, NW lasers based on GaAs have shown great progress, where both pulsed and continuous wave operation have been demonstrated[20–27], and lasing action has been realized not only from photonic modes in horizontal[20,21] and vertical cavities[23], but also from hybrid plasmonic modes in NWs assembled onto flat metallic substrates[26,27]. However, the latter of these, generally known as NW plasmon lasers[28–32], exhibit highly diffractive scattering at the NW ends due to the strong field confinements, and thus the launching of SPPs at the metal/dielectric interfaces has remained elusive. To fully exploit the potential of such a localized and coherent light source in on-chip applications, a platform demonstrating NW laser emission coupling to purposefully designed photonic circuitry remains highly desirable[18].

As a substantial step towards this aim, we demonstrate here hybrid devices consisting of a NW laser, operating at room temperature, that can efficiently launch channel plasmon polaritons (CPPs)[33,34] into lithographically designed and wafer-scale fabricated V-groove (VG) plasmonic waveguides[35]. Additionally, by implementing linear full-wave electromagnetic (EM) simulations and a laser rate-equations analysis, we present convincing arguments that the lasing in our NW-VG system relies on a waveguide CPP-like hybrid mode.

Our hybrid device platform comprises a semiconductor NW positioned at the bottom of a gold (Au) VG channel (Fig. 1a). Upon pulsed illumination, the NW partially couples its laser emission into the propagating CPP modes of the VG, which eventually couple out to free-space at the VG end mirrors. Our core-shell-cap GaAs/AlGaAs/GaAs NWs are grown by a self-catalysed vapour-liquid-solid method (VLS) using Molecular Beam Epitaxy (MBE) on a GaAs (111)B substrate (Supplementary Fig.1)[36–38], with the GaAs core acting as the gain medium. The VG structures were obtained by a recently developed wafer-scale fabrication method of such waveguides on a silicon wafer[35]. We simulated the electric field amplitude distribution of the fundamental CPP mode supported by our VGs at a wavelength of 870 nm (Fig. 1b), corresponding to the band-edge emission of GaAs at room-temperature. The CPPs are a special type of SPPs that are polarized (primarily) parallel to the sample surface (electric field lines at the inset of Fig. 1b), that combine strong EM field confinements (modal lateral size and area of 538 nm and 0.067 $\mu m^2$ respectively) with good propagation characteristics (19.96 $\mu m$ at 870 nm)[39,40]. Once transferred to the VG chip, the NWs were repositioned across the Au film and into the VGs by using micro and nanomanipulation techniques (Supplementary Fig.2a-d). Fig. 1c shows an angled view Scanning Electron Microscopy (SEM) image of an assembled device consisting of a 6.8 µm long NW with a diameter of 590 nm ±45 nm, inside a 30 µm long VG. The end-facet geometry and good alignment within the VG axis are observed in the inset image. Details on the NW growth, VG fabrication, positioning and simulations are available in the Methods section.

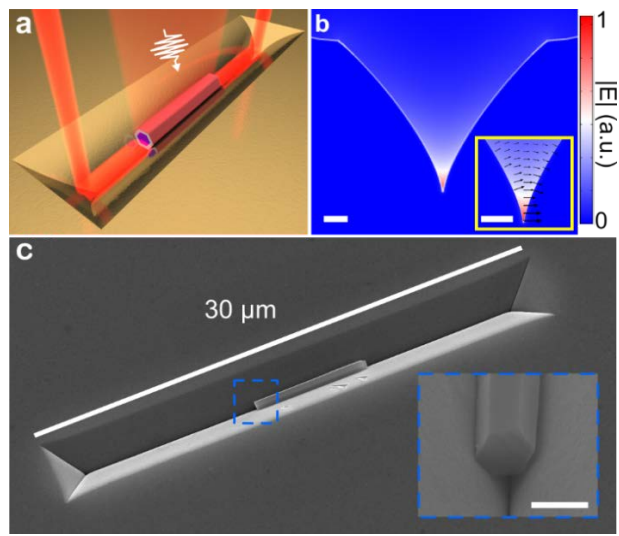

**Figure 1 Hybrid NW-VG platform for an on-chip nanolaser source**. **a**, Schematic illustration of a NW positioned inside a gold VG plasmonic waveguide capable of coupling its lasing emission to the propagating CPP modes upon pulsed optical excitation. **b**, Electric field amplitude |E| profile of the fundamental CPP mode supported by a VG waveguide at 870 nm. The CPP mode is confined at the bottom of the VG with electric field lines mainly transversal to the VG long axis (inset). Scale bars are 300 nm **c**, SEM angled view image of a GaAs/AlGaAs/GaAs NW positioned inside a 30 µm long VG. The inset shows a close-up view of one of the NW end facets (scale bar 550 nm).

To optically characterize our devices, we implemented a homemade micro-photoluminescence optical microscope with a 730 nm pulsed (200 fs, 80 MHz) excitation channel, and two collection channels, including an EMCCD camera for imaging and a fibre collection channel for spectral analysis (see Methods). We first positioned the NWs in the vicinity of the VGs, on top of the Au film, allowing for NW characterization prior to the hybrid device assembly. In essence, this configuration is a reproduction of the NW lasing action of previous works, where similar GaAs-based NWs were randomly deposited onto metallic substrates[26,27]. Figure 2a shows the emission spectra from a NW (385 nm ± 26 nm diameter and 5.5 µm length) lying next to a 30 µm long VG, and excited below (sky-blue) and above (orange) the lasing threshold. The spectra were collected from the right NW facet. Below threshold, we observe a broad spectrum characterized by Fabry-Perot (FP) like oscillations, as expected in such NW cavities[41]. Above the lasing threshold, strong and narrow lasing peaks at ~875 nm and ~868 nm dominate the spectrum. Furthermore, well pronounced interference fringes perpendicular to the NW axis are observed in the EMCCD image taken above the threshold condition (Fig. 2b), manifesting spatial coherence in the emission from the NW facets[42].

Let us consider a schematic of the SPP excitation and scattering in our NW-VG configuration (Fig. 2c) for better understanding of the various observed out-coupled signals (Fig. 2b). A NW placed alongside a VG can couple its emission into two distinct channels, one comprising the light scattered directly from the NW into free-space modes, while the second channel consists of SPPs excited at the Au/air interface (red arrows in Fig.2c). We can identify their signature, since SPPs can scatter out into free-space when encountering defects along their path, in this case, the edges of the VG as observed by the multiple emission spots in Fig. 2b under the transversal polarization collection with respect to the VG long axis (the case for parallel polarization can be found in the Supplementary Fig.3). In passing, we note that this direct evidence of the SPP excitation has often been overlooked when studying such NW lasers on defect free metal films, where only the emission arising from the NW facets has been analysed[26–28,31].

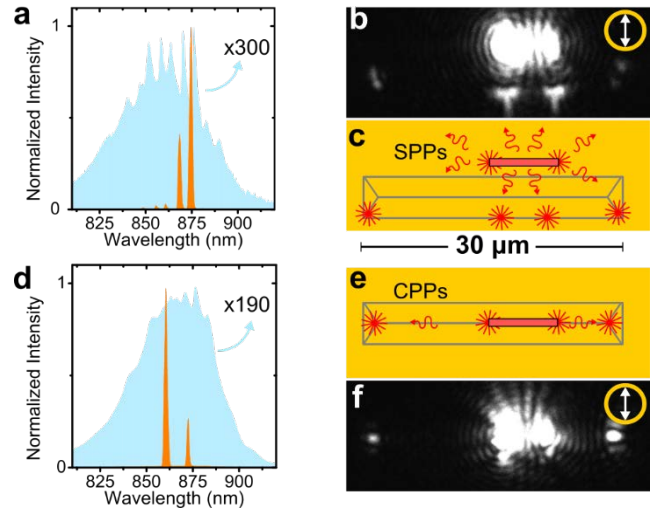

**Figure 2. Optical characterization of a NW laser on a gold film and inside a VG**. **a**, Normalized emission spectra for a NW placed on top of a Au film, below (blue-sky) and above (orange) the lasing threshold. **b**, EMCCD image for the NW excited above the lasing condition, under transversal polarization collection (white arrow), showing interference fringes characteristic of lasing emission. **c**, Schematic illustration of a NW next to a VG. The emission from the NW can either out-couple from the NW facets into free-space or couple its emission into SPPs supported by the Au film/air interface, which subsequently out-scatter from the VG edges. **d**, Normalized emission spectra for the same NW positioned inside the VG, below (blue-sky) and above (orange) the lasing threshold. **e**, Schematic illustration of the NW-VG configuration, where the emission out-couples into free-space at the NW facets, or launches CPPs supported by the VG, which eventually out-scatter at the VG ends. **f**, EMMCD image of the NW-VG device in the lasing regime under transversal polarization collection (white arrow). Two distinct emission spots appear at the VG ends, which correspond to the CPP-coupled emission.

Emission spectra changes significantly (Fig. 2d) once we position the NW at the bottom of the VG as schematically illustrated in Fig. 2e. In this configuration, the emission from the NW can out-couple to free-space via the NW facets, or launch CPPs supported by the VG (red arrows in Fig. 2e), which propagate and out-couple from the two VG ends. Changes in the mode spacing and spectral positions of narrow laser lines (when moving the NW inside the VG), should be explained by considering the available modes in the new hybrid geometry. At the same time, the enhancement of the contribution at longer wavelengths in the emission spectrum, already indicates coupling of the NW emission to CPP modes, whose propagation loss (absorption) decreases rapidly for longer wavelengths[39]. The EMCCD image (Fig. 2f) also clearly demonstrates the CPP excitation from the NW laser, depicted by the two isolated bright emission spots corresponding to the positions of the VG end mirrors. Again, the EMCCD image was acquired above the lasing threshold condition and thus interference fringes are observed.

Next we have a closer look into the polarization dependent emission properties of a second NW-VG device, for which the NW was centred along the VG axis, to provide convincing evidence that we are witnessing lasing generation and coupling to the CPPs supported by our waveguides. We present EMCCD images for the collection with polarization parallel and transversal to the VG main axis (Fig. 3a&c respectively), together with a SEM image of the device to visualize the relative positions of the out-coupled signals (Fig. 3b). The NW (376 nm ± 35 nm in diameter and 6.4 µm long) is symmetrically centred inside the VG, both across and along the VG axis. We observe an evident anisotropy in the emission out-coupled from the VG ends for the transversal polarization configuration (Fig. 3c), providing additional support to the finding that the NW-VG device is able to efficiently launch CPPs[35]. To quantify the polarization dependence, we turn to the spectra collected from four different positions along the VG axis, namely from the NW end facets and the VG ends (Fig. 3d). The spectra are colour coded according to the highlighted dashed circles in Fig. 3a&c (parallel and transversal polarization in magenta and orange respectively). We extract the degree of linear polarization for the lasing peak at ~875 nm, given by $DOLP = (I_\perp - I_\parallel)/(I_\perp + I_\parallel)$, where $I_\perp$ and $I_\parallel$ correspond to the peak intensity for the transversal and parallel polarization collection respectively. At the NW facets, the emission does not exhibit a strong polarization preference, exhibiting $DOLP$ values of 0.21 and 0.29 for the upper and lower NW ends respectively. On the other hand, the situation at the VG ends is drastically different, with large $DOLP$ values of 0.89 and 0.97 for the upper and lower VG ends respectively. These results confirm the excitation of CPPs from the laser emission coupling in our NW-VG platform, given the strong anisotropy in the transversal polarization contributions at the VG ends, in agreement with the CPP characteristics (electric field lines in Fig. 1b).

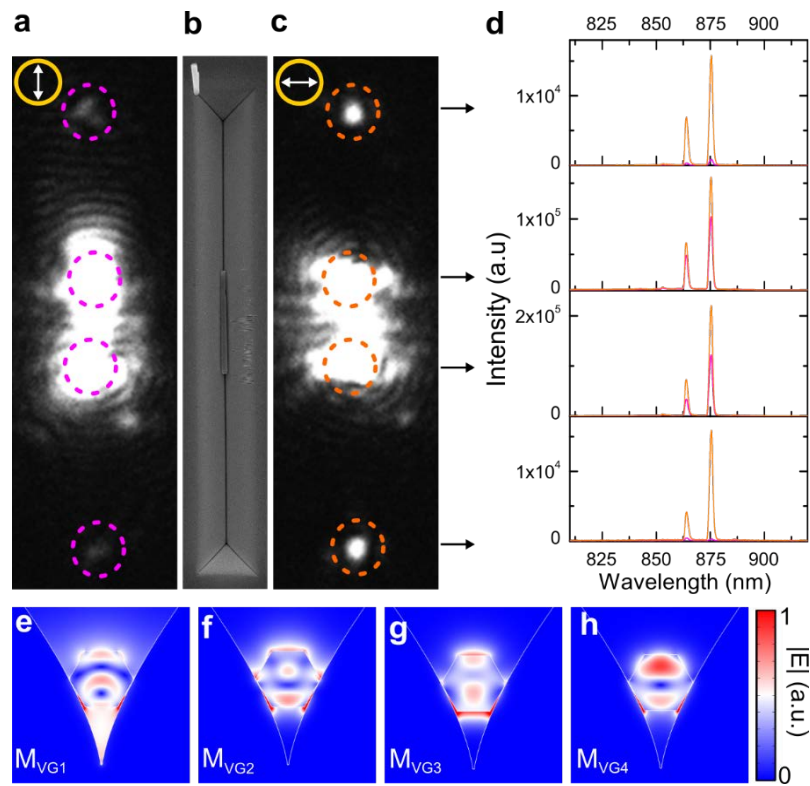

**Figure 3 Evidence of CPP coupling in the NW-VG device**. **a**&**c**, EMCCD images of the NW-VG device in the lasing regime for a collection polarization parallel (**a**) and transversal (**c**) to the VG long axis. **b**, Top view SEM image of the device. **d**, Emission spectra collected from the VG ends and NW end facets for a collection polarization parallel (magenta) and transversal (orange) to the VG axis. The emission at both VG ends exhibits a strong anisotropy in the transversal collection, with a degree of linear polarization ($DOLP$) of 0.89 and 0.97 for the upper and lower VG ends respectively. On the other hand, at the NW ends we extract $DOLP$ values of 0.21 and 0.29 for the upper and lower NW facets respectively. Stronger signals arising from the VG ends for the polarization transversal to the VG are in agreement with the characteristic polarization dependence of the CPPs. **e**-**h**, Simulated |E|-field mode amplitude profiles of the NW-VG system ($M_{VG1-4}$). $M_{VG1}$ is a CPP-like mode while $M_{VG2-4}$ mainly present photonic confinements within the NW geometry.

To understand the behaviour of our hybrid NW-VG system, we carried out 2D and 3D EM simulations in COMSOL Multiphysics (a commercially available implementation of the finite element method, see details in the Supplementary Information and Methods). First, we determined the propagating EM modes supported by the hybrid NW-VG configuration. Several modes were found, many of which have the EM fields mainly confined inside the NW, given its ability to support photonic-like modes[20]. However, only four EM propagating modes are compatible with lasing action, as the other modes exhibit propagation lengths that are shorter than the length of the NWs. We present the E-field profiles of these four modes (Fig. 3e-h), which we labelled as $M_{VG1}$-$M_{VG4}$. Details on the mode characteristics are found in the Supplementary Table 1. The $M_{VG1}$ in particular exhibits a very good mode overlap[43] with the bare CPP mode supported by the VG (25.5% for $M_{VG1}$, versus <0.2% for $M_{VG2}$-$M_{VG4}$). While this can be a fair indication of the mode-matching between the NW-VG modes and the bare CPP mode, the mode overlaps do not account for the physical effects occurring at the NW facets (i.e., back-reflections or scattering).

Instead, to quantify the transfer of energy into the sub-wavelength confined propagating modes, we carried out rigorous 3D FEM simulations to calculate the fraction of energy that channels into the CPP mode compared to the total energy exiting the NW facet, which we term here as transfer efficiency ($\xi$). Note that this definition excludes the fraction of energy that provides the cavity feedback (i.e., the reflection coefficient). We found that $M_{VG1}$ presents the largest value of $\xi$ (23.8%), at least a factor of 15 larger compared to $M_{VG4}$ ($\xi$=1.6%), while the other two modes exhibit negligible $\xi$ values (<0.01%).

Experimentally, we estimate the transfer efficiency from the measurements presented in Fig.3d, quantifying the contributions of the strongest lasing peak (875 nm) collected from one VG end and comparing it to the total emission de-coupled from the corresponding NW facet (free-space scattering and the VG end contribution). We obtain values of 9.3% and 7.1% for the upper and lower halves of the device respectively. Note that these experimental values represent a lower bound since we used the simulated propagation length of the CPP mode (19.96 µm) to compensate the intensities measured from the VG ends, and we did not consider the fraction of energy lost from the CPP to free-space energy conversion at the VG end mirrors.

The latter of these, together with experimental imperfections not present in the theory design, can explain the discrepancy with the simulated values. With an experimental transfer efficiency of nearly 10%, our device stands very well against previous realizations when considering the transfer of energy from an on-chip laser to a sub-wavelength confined propagating mode[7,44,45]. It is difficult to compare against previous NW plasmon lasers (i.e., NWs on metal films), since the SPP transfer signature has remained elusive (i.e., emission only collected from the scattering at the NW facets)[26–28,31]. Our numerical and experimental results suggest that the CPP-like mode supported by the NW-VG system ($M_{VG1}$) enables the lasing action observed in our experiments, since the emission contributions observed at the VG are compatible with the simulated values for $M_{VG1}$.

To demonstrate that our hybrid NW-VG device exhibits lasing action and not just amplified spontaneous emission (ASE)[46,47], we provide here detailed results on the emission behaviour as a function of the incident pump power for the device presented in Fig. 3. The peak intensity of the dominating emission peak (~875 nm), as obtained from the spectra collected from the right VG end (transversal polarization), is related to the average pump power measured before our beam splitter and objective (Fig. 4a). This is typically referred as a Power in-Power out ($P_{in}$-$P_{out}$) plot[48]. The $P_{in}$-$P_{out}$ is characterized by three main regimes. In the first linear regime (lower values of pump power) the spontaneous emission (SE) of the gain medium governs the emission properties of the structure. This regime is followed by a superlinear increase in intensity and a linewidth narrowing, which corresponds to the case in which now ASE dominates the emission characteristics. Finally, for larger values of pump power, the system enters a second linear regime, which is the lasing regime. We stress here the importance of presenting lasing data on a log-log scale[47,48], since a system entering only into the ASE regime will also exhibit a kink in a linear-linear plot (superlinear increase of intensity) and linewidth narrowing, two observations which can mislead the claim of lasing action in active photonic structures[46].

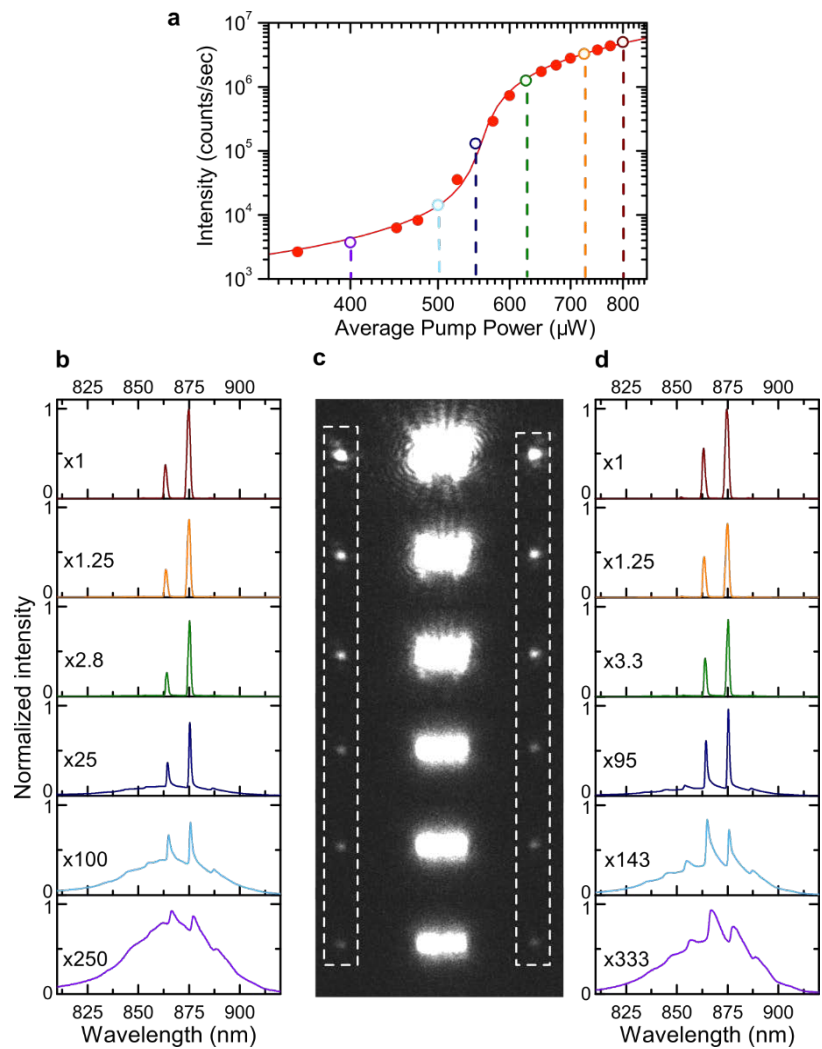

**Figure 4. Lasing sequence of a NW-VG device. a**, $P_{in}$-$P_{out}$ plot of the peak intensity as a function of average pump power (μW) for the dominating lasing peak of a NW-VG device. The signal was collected from the right VG end. The solid red line is a fit using laser rate-equations analysis, which yielded $β=5.5 \times 10^{-4}$ and $R=0.44$. **b&d**, Normalized emission spectra collected from the left and right VG ends respectively, where the spectra are colour coded accordingly to the data points in the $P_{in}$-$P_{out}$ plot. **c**, EMCCD images of the NW-VG device at the respective pump excitation power of each coloured data point. Bright spots can be appreciated at both VG ends, from which the spectra were collected. Interference fringes can be observed in the images associated to the green data point and beyond (625 µW), corresponding to the spatial coherence from the NW facet´s emission in the established lasing regime.

Spectra of the emission from both VG ends (Figs. 4b and 4d) is recorded along with acquiring the associated EMCCD images (Fig. 4c) corresponding to each of the coloured data points in Fig. 4a. At low excitation power (below 450 µW) the system is clearly in the SE regime, with spectra characterized by a broad emission with FP oscillations. The EMCCD image exhibits a bright emission from the NW body and two dim spots at the VG ends (CPP-coupled SE). The sky-blue data point (500 µW) marks the transition into the ASE regime. At the mid-point of this regime (dark-blue data at 550 µW), the spectra are characterized by two narrow and intense peaks. However, if one looks at the EMCCD images, it is not until we surpass the mid-point of the S-shape in the $P_{in}$-$P_{out}$ curve (e.g. green data at 625 µW), that we observe interference fringes characteristic of transitioning into the well-developed lasing regime. From this point forward, the spectra are characterized by the strongly dominating lasing peaks and pinning of the spontaneous emission fraction. The lasing peak intensity follows a linear dependence as a function of excitation pump power (orange and maroon data points). The EMCCD images were acquired under the same exposure parameters and thus have been post-processed (for brightness) to clearly distinguish the CPP related out-coupling spots at the VG ends (raw images available in Supplementary Figure 4).

The solid red line in Fig. 4a is a fit to the experimental data using a laser rate-equations analysis. In this case, we take the CPP-like $M_{VG1}$ mode as the input EM field (details in the Supplementary Information). There are two fitting parameters in our modelling, namely the coupling factor of spontaneous emission into the lasing mode (β-factor), and the reflectivity

(R) of the mode at the NW end facet. For the case depicted in Fig. 4a, we obtained values of $\beta=5.5\times10^{-4}$ and R=0.44, which agrees very well with the one obtained from a 3D linear EM simulation for the $M_{VG1}$ mode (R=0.46, see Supplementary Table 1). It is worth mentioning that when fitting the same experimental data to the other three modes, we obtained similar values for β, but values of R that depart considerably from the ones obtained with the linear simulations (for instance, for $M_{VG4}$ the rate-equations fit yields $\beta=4.5\times10^{-4}$ and R=0.35, while full-wave simulations predict R=0.73). The very good agreement between the reflectivity deduced from our rate-equations analysis and 3D simulations for the CPP-like mode together with the experimental evidence that our system clearly couples its lasing emission into the propagating CPP modes, provide convincing arguments to suggest that we indeed observe lasing action from the $M_{VG1}$ mode supported by the NW-VG architecture.

In this Article, we have demonstrated a plasmonic waveguide-integrated NW laser platform, operating at room temperature and deterministically assembled using micro and nanomanipulation techniques. We showed clear evidence of the lasing transition and the coupling to CPP modes in our hybrid NW-VG devices, exhibiting a remarkable transfer efficiency of the lasing emission into a sub-wavelength mode of nearly 10%. Theoretical support was key to reveal that a CPP-like hybrid plasmonic mode likely enables the lasing emission observed in our system.

Since their introduction, NW lasers have been envisioned as fundamental elements in future on-chip sensing and data communication systems[18]. Our novel NW-VG hybrid devices represent the first realization, to the best of our knowledge, of a NW laser integrated with a wafer-scale lithographically designed waveguide, paving the way for such integrated nanolaser sources in a wide variety of on-chip applications. Furthermore, this platform can be exploited by integrating other high quality NW heterostructures, as those featuring quantum confined active regions to improve the performance of these devices[25,27,49]. We envision that the open-channel nature and EM field confinement of the CPPs will provide the means to build high-sensitivity chemical or bio-sensing platforms with an integrated nanolaser source. Additionally, other elements can be integrated along the VG to build functional photonic circuitry, for example in quantum optics experiments employing single quantum emitters coupled to the CPPs[50,51], or integrating electro-optic elements in high-speed plasmonic modulators for data communication[13,52]. Future directions should focus on the development of electrical injection based waveguide integrated NW lasers, with previous efforts indicating this is feasible[53–55]. The latter combined with integrated modulators and photodetectors could bring us one step closer to optics-less hybrid photonic-plasmonic circuit platforms.

**Methods**

**Nanowire growth**: The nanowires (NWs) used in this study are grown using a DCA solid source Molecular Beam Epitaxy (MBE) system (DCA Instruments) with Gallium (Ga) assisted self-catalysed vapour liquid solid (VLS) technique[36]. Gallium-Arsenide (GaAs) substrates were coated with a Hydrogen silsesquioxane (HSQ) converted $SiO_x$ (1<x<2) layer of 4 nm (±0.5nm)[38]. GaAs NW cores (~200 nm in diameter) are grown at 630 °C with 7 r.p.m. rotation speed under a planar growth rate of 0.48 Å/s and a V/III ratio of 6.35 on a (111)B GaAs substrate. The NW thickness was controlled with two parameters, duration of shell growth and NW density. Upon achieving the core growth of the NWs, a Ga droplet has been consumed under Arsenic (As) flux, and the growth conditions are switched from axial to radial growth by increasing V/III ratio through As flux to 37.3, and decreasing the temperature to 465 °C (Supplementary Fig.1). A GaAs shell is radially grown around the core of GaAs nanowires in order to control the thickness of the nanowires in a systematic way. The GaAs shell thickness have been varied from 55 nm to 115 nm. Following that, nanowires are capped with an $Al_xGa_{1-x}As$ (x=0.25) shell of 25 nm for surface passivation and another 10 nm of GaAs shell to avoid oxidization. The NW density needed to be controlled in order to achieve a homogeneous shell thickness along the NWs with reduced tapering. A moderate NW density was realized for that purpose by modifying the Ga rate and $SiO_x$ layer thickness.

**V-groove fabrication**: A silicon substrate with a 200 nm $SiO_2$ layer is patterned by both UV-lithography and reactive ion etching to define the perimeter of the V-groove (VG) devices (initial width of 3.5 μm and varying lengths)[35]. The VG and termination mirrors are formed by anisotropic wet etching of the exposed silicon in a potassium hydroxide (KOH) bath at 80 °C. The KOH etch yields smooth ⟨111⟩ VG sidewalls and termination mirrors with a fixed inclination of 55° from the surface plane. The remaining SiO2 is removed by etching in a hydrofluoric (HF) acid bath. Tailoring of the V-shape geometry is performed by thermal wet oxidation of the silicon V-grooves at 1150 °C for 9 h, resulting in a 2320 nm thick SiO2 layer at flat sections of the substrate. The VG widths are approximately 3.2 µm after the thermal oxidation step. The gold film is deposited by electron beam evaporation: first a 5 nm layer of chromium to promote adhesion before the 70 nm layer of gold. The gold layer is chosen to be sufficiently thick to eliminate interaction of air−interface plasmons with the SiO2 layer and to also minimize aggregation.

**NW transfer and positioning**: The NWs were transferred from the as-grown substrate to the VG chip with a dry transfer technique, consisting of swiping a piece of clean room tissue (Berkshire) at the as-grown substrate, and subsequently repeating this at the VG area, leaving NWs lying across the chip. For the micromanipulation step, we installed a micromanipulator at a conventional microscope (Olympus BX51), and used glass fibre tips fabricated with an optical fibre puller (Sutter P2000). We monitor in real-time the movements of the NWs across the chips under a long working distance 50x, 0.5NA objective (Olympus) and imaging with a CMOS camera (Pixelink). Supplementary Figure 2 presents snapshots of the positioning of four NW-VG devices, along with SEM images of the assembled devices. For NWs with diameters below 500 nm, the fibre tips were not able to push the NWs to the bottom of the VGs. For those cases, we positioned the NWs initially at the edge of the VGs and later positioned them inside the VGs by using an AFM tip in contact mode (Nanoman, Dimension 3100 Veeco)[50].

**Optical characterization**: The substrate with VGs and NWs were mounted on a 3-axis piezoelectric stage installed on a 2-axis micrometre controlled stage, providing us with short and long range movements across the plane of the sample. Pulsed optical excitation was done with a Ti:Sapphire laser (Coherent) tuned at 730 nm (repetition rate 80 MHz and pulse duration 200 fs). The excitation light was focused on the sample with a 40x, 0.65 NA objective (Edmund). Introducing a cylindrical lens (f=1 m) on the excitation path provided us with an elliptical beam of approximately 18.8 µm$^2$ (diameter along the major and minor axes of 20 µm and 1.2 µm respectively). The focused area was sufficiently large to excite the entire NWs. Before the objective we installed a 50:50 beamsplitter (BS), and a neutral density (ND) filter wheel was placed before the BS to control the excitation power reaching the sample, measured with a power meter after the ND filter. The emission from the NWs was collected through the same objective, and after passing the BS, we placed a 785 nm long pass filter and a linear polarizer. The emission was split into two channels using another 50:50 BS. On one channel we focused the emission into an EMCCD camera (Hamamatsu) providing us with a wide-field view of the emission from the NWs and NW-VG devices. On the second channel we focused the emission into a single-mode optical fibre which was coupled to a spectrometer (Andor Shamrock SR163). The detection at the fibre allowed us to selectively collect the emission from different areas around the excitation spot with a steering mirror. When necessary, to maintain constant acquisition parameters, ND filters were introduced as the emission intensities from the NW facets were stronger than the one collected at the VG ends.

**Theory simulations**: We performed 2-dimensional (2D) numerical simulations of the supported EM of the VG structure and of the NW/VG hybrid structure, by using the finite element method (FEM), implemented by the commercial package Comsol Multiphysics. This tool enables us to calculate the propagating eigenmodes for the considered translationally invariant structures at the operating frequency ($\omega_e=2\pi c/\lambda_e$, with $\lambda_e=870$ nm), and gives access to the corresponding electric field amplitude distributions. The geometry of the VG cross-section profile was generated from numerical process simulations in ATHENA (Silvaco, Inc), and subsequently exported to the Comsol interface[35]. The considered nanowires have a diameter of 370 nm, with a hexagonal cross section. They are situated at the VG-bottom such that two of their adjacent vertices lie at the same height on the metal surface. The material properties were defined as the following: the Au optical response is described by a Drude-Lorentz model, fitted to available experimental data[56,57]. The nanowire is considered uniform, with a refractive index of $n_{NW}=3.6$. Moreover, the ensemble NW/VG is embedded in air, with a refractive index equal to 1.

The resulting modes from the 2D FEM simulations are characterized by an effective mode index $n_{eff}$. By inspection up to $n_{eff}=4$, a total number of nine modes was found. One of them was CPP-like and the eight remaining modes were mainly confined inside the nanowire. The electric field intensity of each of these modes decays along the VG long-axis in an exponential fashion: $I=I_0 \exp(-z/L_p)$, where $z$ is the transversal direction and $L_p$ is the propagation length, given by $L_p=(2\text{Im}(n_{eff}k_0))^{-1}$. We selected those modes whose propagation length is higher than the considered nanowire length of 6.5 µm, yielding the four modes $M_{VG1}$-$M_{VG4}$ presented in the main text. Values of the effective mode indices and of the propagation lengths can be found in the Supplementary Table 1, together with details on the 3D FEM simulations and on the rate-equation approach.

## ASSOCIATED CONTENT

**Supporting Information**

Experimental details on the fabrication and theoretical simulations. This material is available free of charge via the Internet at http://pubs.acs.org.

**Acknowledgements**

E.B.-U. and R.Q. acknowledge financial support from the European Community's Seventh Framework Program under grants ERC-Plasmolight (259196) and QnanoMECA (64790), the Spanish Ministry of Economy and Competitiveness, through the 'Severo Ochoa' Programme for Centres of Excellence in R&D (SEV-2015-0522) and grant FIS2013-46141-P, Fundació Privada Cellex, and the CERCA Programme from the Generalitat de Catalunya. E.B.-U acknowledges support from the FPI fellowship from the Spanish MICIIN, and Universidad de Costa Rica. G.T and A.F.i.M acknowledge funding by SNSF through the NCCR QSIT, and technical support in the MBE by J. B. Leran, H. Potts, F. Matteini and F. Jabeen. J.C., J.B.-A. and F.J.G.-V. acknowledge financial support from the European Union's Seventh Framework Program under grants ERC-PlasmoNanoQuanta (290981) and the Spanish Ministry of Economy and Competitiveness, through the 'Maria de Maeztu' Programme for Units of Excellence in R&D (MDM-2014-0377) and grants MAT2014-53432-C5-5-R and MAT2015-66128-R. C.L.C.S acknowledges support from the Danish Council for Independent Research through grant 12-126601. S.I.B acknowledges financial support for this work from the European Research Council, Grant No. 341054 (PLAQNAP). We thank D. Saxena for fruitful discussions, and M.Svendendahl and L. Novotny for feedback on our manuscript.

**Author information**

**Corresponding Authors**

*E-mail: romain.quidant@icfo.es

*E-mail: esteban.bermudez@icfo.es

## Author contributions

E.B.-U and R.Q. conceived the experiment. G.T. developed the NW growth with input from A.F.i.M. C.L.C.S. fabricated the VGs. E.B.-U performed the hybrid device assembly, optical characterization and data analysis. J.C. Carried out the 2D and 3D simulations in COMSOL. J.B.-A. implemented the laser rate-equations model. J.C, J.B.-A and F.J.G-V analysed the theory results. E.B.-U wrote the first draft of the manuscript with further input from all authors. S.B., A.F.i.M, F.J.G-V and R.Q supervised the work.

† These authors contributed equally.

## Notes

The authors declare no competing financial interests.